\documentclass[aps,prd,onecolumn,groupedaddress,showpacs,nofootinbib,amssymb]{revtex4}
\usepackage[latin1]{inputenc}
\usepackage{graphicx}
\usepackage[english]{babel}
\usepackage{amsmath}
\usepackage{amssymb}
\usepackage{amsfonts}
\usepackage{color}

\allowdisplaybreaks[4] 

\begin{document}

\newcommand{\be}{\begin{equation}}
\newcommand{\ee}{\end{equation}}
\newcommand{\bea}{\begin{eqnarray}}
\newcommand{\eea}{\end{eqnarray}}
\newcommand{\beaa}{\begin{eqnarray*}}
\newcommand{\eeaa}{\end{eqnarray*}}
\newcommand{\Lhat}{\widehat{\mathcal{L}}}
\newcommand{\nn}{\nonumber \\}
\newcommand{\e}{\mathrm{e}}
\newcommand{\tr}{\mathrm{tr}\,}

\tolerance=5000

\title{Critical gravity with a scalar field in four dimensions}

\author{Kyosuke Hirochi$^1$ and 
Shin'ichi Nojiri$^{1,2}$}

\affiliation{
$^1$ Department of Physics, Nagoya University, Nagoya
464-8602, Japan \\
$^2$ Kobayashi-Maskawa Institute for the Origin of Particles and
the Universe, Nagoya University, Nagoya 464-8602, Japan 
}

%\author{Kyosuke Hirochi}
\date{}
\def\theequation{\thesection.\arabic{equation}}
%\makeatletter
%\@addtoreset{equation}{section}
%\makeatother

%\begin{document}

%\maketitle

\begin{abstract}

We consider the critical gravity theory with a scalar field in four dimensions. 
We find that this theory has the solution corresponding to the de Sitter (dS), 
anti-de Sitter (AdS), and Minkowski 
background depending on whether the action includes the cosmological term or not. 
The Minkowski background is the solution which cannot be obtained in the model 
without a scalar field.
At the critical point, we show that the Abbott-Deser (AD) mass of the Schwarzschild-de Sitter (SdS) 
black hole and the energy for the massless graviton vanish, whose situation is not changed 
from the model without a scalar field.

\end{abstract}

%\pacs{95.36.+x, 98.80.Cq}

\maketitle

%\newpage
%\tableofcontents

\section{Introduction \label{SecI}}

As well-known, Einstein's theory of the gravity describes the nature very well 
in the classical level. This theory does not, however, work as a quantum theory 
due to the non-renormalizablity.
Therefore as a quantum theory of gravity, we need to find a model beyond Einstein's theory.

Recently, a candidate of the quantum gravity in four dimensions was proposed in \cite{1}, 
which could be renormalizable and unitary by adding a term including the squared Weyl tensor 
to the Einstein-Hilbert (EH) action with a cosmological term. 
Usually the theories including the terms given by the square of the curvatures have the massive spin 2 mode 
and the massive scalar (spin 0) mode in addition to the massless graviton. 
In a special case that the curvature squared term is given by the square of the Weyl tensor, 
however, the massive scalar mode does not appear, which can be shown by choosing an appropriate gauge condition. 
Furthermore, by a special choice of the parameters, the massive spin 2 mode becomes massless and the mode 
degenerates with the massless graviton. 
By this special choice of the parameters, which is called as a critical point, 
there appears a mode which behaves as a logarithmic function of the distance. 
The appearance occurs due to the degeneracy of the massless spin 2 mode, 
which is described by the fourth order 
differential equation.
The logarithmic mode has positive energy at the critical point but a combination of the massless 
and logarithmic mode could generate negative energy as shown in \cite{2}. 
Thus by the requirement of the unitarity, the logarithmic mode must be truncated 
by an appropriate boundary condition.

In this paper, we consider a model of the critical gravity in four dimensions with a scalar field. 
The model includes a non-minimal coupling between the scalar field and the scalar curvature as in the 
Brans-Dicke model. 
We find that this theory has solutions describing the de Sitter (dS), 
anti-de Sitter (AdS), and Minkowski backgrounds. 
We also investigate the energies of the propagating modes and the Abbott-Deser (AD) mass 
of the Schwarzschild-de Sitter (SdS) black hole in each of the backgrounds 
and we find that they vanish at the critical point.

\section{Equations of motion in critical gravity with a scalar field \label{SecII}}

We consider a model of the critical gravity with a scalar field, whose action is given by
\be
\label{1}
S=\int d^4x \sqrt{-g} \left\{ \frac{1}{2\kappa^2}\left[ R - 2\Lambda_0
+\alpha \left(R^2_{\mu\nu}-\frac{1}{3}R^2 \right) \right] 
 -\frac{1}{2}(\nabla_\mu\phi)^2-\frac{1}{2}m^2\phi^2
+\gamma R\phi^2-\frac{1}{4!}\xi \phi^4\right\}\, ,
\ee
where $\alpha$, $\gamma$, and $\xi$ are coupling constants. 
We now call $\Lambda_0$ as ``bare'' cosmological constant. 
It is known that this kind of action is power-counting renormalizable \cite{Stelle} 
(for a review, see \cite{Buchbinder:1992rb}). However, being higher-derivative one, such 
theory is known to have the unitarity problem which is not solved yet. 

The equations of motion given by the variation of $\phi$ and $g_{\mu\nu}$ 
have the following forms, respectively:
\bea
\label{2}
&& \left(\nabla^2-m^2+2\gamma R\right)\phi=\frac{1}{3!}\xi\phi^3\, ,\\
\label{3}
&& \mathcal{G}_{\mu\nu}+E_{\mu\nu}+\Phi_{\mu\nu}=0\, ,
\eea
where $\mathcal{G}_{\mu\nu}$ is given by the variation of the Einstein-Hilbert term 
with a bare cosmological constant. 
On the other hand, $E_{\mu\nu}$ and ${\Phi}_{\mu\nu}$ express the contributions 
from the term including the 
square of the Weyl tensor and the terms including the scalar field, respectively. 
These contributions are explicitly expressed as 
\begin{eqnarray}
\label{4}
\mathcal{G}_{\mu\nu}&=&R_{\mu\nu}-\frac{1}{2}Rg_{\mu\nu}+\Lambda_0 g_{\mu\nu}\, ,\\
\label{5}
E_{\mu\nu}&=&\alpha\left[2R_{\mu\rho}R^\rho_\nu-\frac{1}{2}R^2_{\rho\sigma}g_{\mu\nu}
 -\frac{2}{3}R \left( R_{\mu\nu}-\frac{1}{4}Rg_{\mu\nu} \right) 
+\nabla^2 \left( R_{\mu\nu}-\frac{1}{6}Rg_{\mu\nu} \right) \right. \nn
&& \left. + \frac{2}{3}\nabla_\mu\nabla_\nu R 
 -2\nabla_\rho\nabla_{(\mu}R_{\nu)}^\rho\right]\, , \\
\label{6}
{\Phi}_{\mu\nu}&=&2\kappa^2 \left[\frac{1}{2}g_{\mu\nu}
\left\{ \frac{1}{2}(\nabla_\rho\phi)^2+\gamma(2\nabla^2-R)\phi^2
+\frac{1}{2}m^2\phi^2+\frac{1}{4!}\xi\phi^4 \right\} 
 -\frac{1}{2}\nabla_\mu\phi\nabla_\nu\phi
+\gamma(R_{\mu\nu}-\nabla_\mu\nabla_\nu)\phi^2 \right]\, .
\end{eqnarray}

By assuming $\phi$ is a constant, $\phi=\phi_c$ in Eqs.~(\ref{2}) and (\ref{3}), 
we obtain the vacuum solution as follows,
\begin{equation}
\label{7}
R_{\mu\nu}=\Lambda\, ,\quad 
R=4\Lambda\, ,\quad 
R_{\mu\nu\rho\sigma}=\frac{\Lambda}{3}(g_{\mu\rho}g_{\nu\sigma}-g_{\mu\sigma}g_{\nu\rho})\, ,
\end{equation}
where 
\begin{equation}
\label{8}
\Lambda\equiv\Lambda_0+\frac{1}{4}bm^2\, ,\qquad 
b\equiv\kappa^2\phi_c^2\, .
\end{equation}
We should note that we have used the relation between the coupling constants, 
\begin{equation}
\label{9}
\xi\phi^2_c=6(8\gamma\Lambda-m^2)\, .
\end{equation}
which is derived from (\ref{2}) and we have eliminated the coupling constant $\xi$ 
in (\ref{3}). 
Eq.~(\ref{8}) tells that the solution corresponding to the Minkowski background 
is realized when
\begin{equation}
\label{10}
\Lambda_0=-\frac{1}{4}bm^2\, .
\end{equation}
If $\Lambda_0>-\frac{1}{4}bm^2$ $\left(\Lambda_0<-\frac{1}{4}bm^2\right)$, we 
obtain the solution corresponding to dS (AdS) background. 
It is clear that the solution corresponding to the Minkowski background in (\ref{10}) 
is generated due to the existence of the scalar field.

\section{Perturbations around the background and linearized equations \label{SecIII}}

In the last section, we have shown that Eqs.~(\ref{2}) and (\ref{3}) have three 
kinds of the background solutions corresponding to the (A)dS and Minkowski space-times. 
In this section, we consider the perturbations from the backgrounds and derive the linearized equation. 
After that, we will show that there is a critical point where the massive spin 2 mode 
reduces to massless one.

The perturbations from the backgrounds are given by
\begin{equation}
\label{3-1}
g_{\mu\nu}\rightarrow g_{\mu\nu}+h_{\mu\nu}\, ,
\quad \phi_c\rightarrow\phi_c+\phi\, .
\end{equation}
We find the linearized equations for $\phi$ and $h_{\mu\nu}$ have the following forms:
\begin{eqnarray}
\label{3-2}
0&=& \left( \nabla^2+2m^2-16\gamma\Lambda \right) \phi 
+ 2\gamma\phi_c R^L \, ,\\
\label{3-3}
0&=&\mathcal{G}_{\mu\nu}^L+E_{\mu\nu}^L+\Phi_{\mu\nu}^L \nonumber \\
&=& \left\{ 1+2b\gamma+\alpha \left( \nabla^2-\frac{4\Lambda}{3} \right) \right\}
\tilde{\mathcal{G}}_{\mu\nu}^L+\frac{\alpha}{3}\left( g_{\mu\nu}\nabla^2
 -\nabla_\mu\nabla_\nu-\Lambda g_{\mu\nu} \right)R^L \nn
&& +\frac{4b\gamma}{\phi_c}\left (g_{\mu\nu}\nabla^2
 -\nabla_\mu\nabla_\nu+\Lambda g_{\mu\nu} \right)\phi\, ,
\end{eqnarray}
where
\begin{eqnarray}
\label{3-4}
\mathcal{G}_{\mu\nu}^L&=&R_{\mu\nu}^L-\frac{1}{2}R^Lg_{\mu\nu}
-\Lambda h_{\mu\nu}+(\Lambda_0-\Lambda)h_{\mu\nu} \nonumber\\
&=&\tilde{\mathcal{G}}_{\mu\nu}^L-\frac{1}{4}bm^2h_{\mu\nu}\, ,\\
\label{3-5}
E^L_{\mu\nu}&=&\alpha\left\{ \left(\nabla^2-\frac{4\Lambda}{3}\right)
\tilde{\mathcal{G}}_{\mu\nu}^L+\frac{1}{3} \left( g_{\mu\nu}\nabla^2
 -\nabla_\mu\nabla_\nu-\Lambda g_{\mu\nu} \right)R^L \right\}\, ,\\
\label{3-6}
\Phi^L_{\mu\nu}&=&2b\gamma\tilde{\mathcal{G}}^L_{\mu\nu}
+\frac{1}{4}bm^2h_{\mu\nu}+\frac{4b\gamma}{\phi_c} \left(g_{\mu\nu}\nabla^2
 -\nabla_\mu\nabla_\nu+\Lambda g_{\mu\nu} \right)\phi\, .
\end{eqnarray}
Here we denote the linearized forms of $R_{\mu\nu}$ and $R$ 
by $R^L_{\mu\nu}$ and $R^L$,
\bea
\label{3-7}
&& R^L_{\mu\nu}=\nabla^\lambda\nabla_{(\mu}h_{\nu)\lambda}
-\frac{1}{2}\nabla^2h_{\mu\nu}-\frac{1}{2}\nabla_\mu\nabla_\nu h\, ,\\
&& \label{3-8}
R^L=\nabla^\mu\nabla^\nu h_{\mu\nu}-\nabla^2h-\Lambda h\, ,
\eea
and $\tilde{\mathcal{G}}_{\mu\nu}^L$ is the covariantly conserved part 
$\left( \nabla^\mu \tilde{\mathcal{G}}_{\mu\nu}^L =0 \right)$
of $\mathcal{G}_{\mu\nu}^L$ ($\mathcal{G}_{\mu\nu}^L$ itself is 
not covariantly conserved, $\nabla^\mu \mathcal{G}_{\mu\nu}^L\neq 0$).
By the parentheses $(\ )$ for the indexes, we means the symmetrization with 
respect to the indexes. 
We also used (\ref{9}) in order to eliminate $\xi$ from the linearized equations.

It could be convenient to impose the following gauge condition, 
\begin{equation}
\label{3-9}
\nabla^\mu h _{\mu\nu}=\nabla_\nu h\, .
\end{equation}
In this gauge, (\ref{3-2}) has the following form:
\begin{equation}
\label{3-10}
\left( \nabla^2+2m^2-16\gamma\Lambda \right)\phi-2\Lambda\gamma\phi_ch=0\, ,
\end{equation}
and we can find the equation for the trace of $h_{\mu\nu}$, $h\equiv h_\mu^{\ \mu}$, 
as follows, 
\be
\label{3-11}
0 = g^{\mu\nu} \left(\mathcal{G}^L_{\mu\nu}+E^L_{\mu\nu}+\Phi^L_{\mu\nu} \right) 
= \Lambda(1+2b\gamma)h+\frac{4b\gamma}{\phi_c}(3\nabla^2+4\Lambda)\phi\, .
\ee
Eliminating $\nabla^2\phi$ by substituting (\ref{3-10}) into (\ref{3-11}), 
we obtain the relation between $\phi$ and $h$ as follows,
\begin{equation}
\label{3-12}
\Lambda(24b\gamma^2+2b\gamma+1)h+\frac{8b\gamma}{\phi_c}(24\gamma\Lambda-3m^2
+2\Lambda)\phi=0\, .
\end{equation}
In (\ref{3-12}), if we choose the parameters so that the coefficient of the $h$ 
vanishes, that is, if $24b\gamma^2+2b\gamma+1=0$, $\phi$ also vanishes. 
Furthermore if $\phi=0$, Eq.~(\ref{3-10}) tells that the scalar mode of the graviton $h$ 
also vanishes, $h=0$. 
The condition that the coefficient of the $h$ vanishes in (\ref{3-12}) can be solved with 
respect to $\gamma$, as follows, 
\begin{equation}
\label{3-13}
\gamma=\gamma_\pm\equiv\frac{1}{24}\left\{-1\pm\sqrt{1-\frac{24}{b}}\right\}\, .
\end{equation}
In order that $\gamma_\pm$ could be real, the parameter $b$ should be chosen so that 
\begin{equation}
\label{3-14}
b\geq 24\, .
\end{equation}
We should note that $b$ is positive by the definition in (\ref{8}). 
We should also note that $\gamma_\pm$ is always negative for (\ref{3-14}).

By eliminating scalar graviton, 
we obtain the fourth order partial differential equation for $h_{\mu\nu}$ as follows,
\begin{equation}
\label{fourth}
0=\left(\nabla^2-\frac{2}{3}\Lambda\right)\left\{\nabla^2-\frac{4}{3}\Lambda
+\frac{1}{\alpha}\left(1+2b\gamma_\pm\right)\right\}h_{\mu\nu}\, .
\end{equation}
This equation describes a massless graviton and a massive one, which satisfy the 
following equations, respectively,
\be
\label{17}
\left( \nabla^2-\frac{2\Lambda}{3} \right)h^{(m)}_{\mu\nu}=0\, ,\quad 
\left( \nabla^2-\frac{2\Lambda}{3}-M^2 \right)h^{(M)}_{\mu\nu}=0\, ,
\ee
where
\begin{equation}
\label{Msquared}
M^2\equiv\frac{2\Lambda}{3}-\frac{1}{\alpha}(1+2b\gamma_\pm)\, .
\end{equation}
Since $M^2$ is the mass of the massive graviton, we require $M^2$ should be 
positive semi-definite, $M^2\geq 0$.

We can find a critical point where the massive graviton reduces to massless. 
By using (\ref{Msquared}), we find that the critical point is given by 
\begin{equation}
\label{3-19}
\alpha=\frac{3}{2\Lambda}\left\{1+2b\gamma_\pm\right\}\, .
\end{equation}
At the critical point, the fourth order differential equation for 
$h_{\mu\nu}$ (\ref{fourth}) becomes
\begin{equation}
\left( \nabla^2-\frac{2\Lambda}{3} \right)^2 h_{\mu\nu}=0\, .
\end{equation}
This equation dose not describe only the massless mode but also the logarithmic mode, 
which satisifies the following equations:
\begin{equation}
\left( \nabla^2-\frac{2\Lambda}{3} \right)^2h_{\mu\nu}^{\mathrm{(log)}}=0\, ,\quad
\left( \nabla^2-\frac{2\Lambda}{3} \right)h_{\mu\nu}^{\mathrm{(log)}} =h_{\mu\nu}^{(m)}\, .
\end{equation}
Here $h_{\mu\nu}^{(m)}$ satisfies the first equation in (\ref{17}). 
The logarithmic mode is discussed in \cite{2,3,4}.

We should note that in the perturbation in the Minkowski background, 
there appear several differences with those in the (A)dS background. 
Since $\Lambda=0$ in the Minkowski background, the equation (\ref{3-12}) 
for the trace of $h_{\mu\nu}$ 
after choosing the gauge condition (\ref{3-9}) becomes
\begin{equation}
\phi=0\, .
\end{equation} 
So in contrast with the case of the (A)dS case, there does not appear any condition 
for eliminating the scalar graviton $h$ from (\ref{3-10}). 
In order to eliminate $h$, we can use the residual gauge transformations 
$\delta x^\mu=\partial^\mu\eta$ (see \cite{5}).
Then the fourth order differential equation (\ref{fourth}) 
for $h_{\mu\nu}$ becomes
\begin{equation}
(\partial^2-\tilde M^2)\partial^2 h_{\mu\nu}=0\, ,
\quad \tilde M^2\equiv-\frac{1+2b\gamma}{\alpha}\, .
\end{equation}
In the massless limit $\tilde M\rightarrow 0$, we find the critical point is given by
\begin{equation}
\label{critical2}
\gamma=-\frac{1}{2b}\, .
\end{equation}

%%%%%%%%%%%%%%%%%%%%%%%%%%%%%

Here we have some comments on the coupling constants. 
We can find that in any backgrounds, $\gamma$ must be negative at the critical point 
(note that at the non-critical point, $\gamma$ can be an arbitrary value in 
the Minkowski background). 
Furthermore $\alpha$ is also arbitrary in the Minkowski background, 
while, in the (A)dS background $\alpha$ is related with other couplings 
as in (\ref{3-19}) at the critical point.   

In the following, we consider the Hamiltonian formalism in order to obtain the on shell energies 
for each of the propagating modes. 
(Here we only consider the case of (A)dS background in detail, but any results in 
the Minkowski one can be easily obtained by putting $\Lambda=0$ and $g_{\mu\nu}=\eta_{\mu\nu}$).

First, we introduce the quadratic action which is derived from 
the linearized equation for $h_{\mu\nu}$. 
Using the gauge condition (\ref{3-9}) and choosing $\gamma=\gamma_\pm$ as in (\ref{3-13}) 
in order to eliminate $h$ and $\phi$, we obtain
\begin{eqnarray}
I_2&=&-\frac{1}{2\kappa^2}\int d^4 x \sqrt{-g} h^{\mu\nu}(\mathcal{G}_{\mu\nu}^L
+E_{\mu\nu}^L+\Phi_{\mu\nu}^L) \nonumber\\
&=&-\frac{1}{2\kappa^2}\int d^4x \sqrt{-g} \left\{(1-\frac{4}{3}\alpha\Lambda
+2b\gamma_\pm)\frac{\Lambda}{3}h^{\mu\nu}h_{\mu\nu} 
+ \frac{1}{2}(1-2\alpha\Lambda+2b\gamma_\pm)(\nabla_\lambda h_{\mu\nu})^2
 -\frac{\alpha}{2}(\nabla^2h_{\mu\nu})^2\right\}\, .
\end{eqnarray}
By using the Ostrogradsky method, we define the canonical momenta as follows,
\begin{eqnarray}
\pi_{(1)}^{\mu\nu}&=&\frac{\delta L_2}{\delta\dot h_{\mu\nu}} 
-\nabla_0\left(\frac{\delta L_2}{\delta(d(\nabla_0h_{\mu\nu})/dt)} \right) 
= -\frac{\sqrt{-g}}{2\kappa^2}\nabla^0\left\{(1-2\alpha\Lambda
+2b\gamma_\pm)h^{\mu\nu}+\alpha\nabla^2h^{\mu\nu}\right\},\nonumber \\
\pi^{\mu\nu}_{(2)}&=&\frac{\delta L_2}{\delta(d(\nabla_0h_{\mu\nu})/dt)}
=\frac{\sqrt{-g}}{2\kappa^2}\alpha g^{00}\nabla^2h^{\mu\nu}\, .
\end{eqnarray}
Since the Lagrangian does not depend on time, the Hamiltonian does not, either. 
Thus we obtain the following Hamiltonian
\begin{equation}
\label{Hamiltonian}
H= - \frac{1}{2\kappa^2T}\int d^4 x \sqrt{-g} \left\{ 
\left(1-2\alpha\Lambda+2b\gamma_\pm \right )\nabla^0 h^{\mu\nu}\dot h_{\mu\nu}
+2\alpha\frac{\partial}{\partial t}\left(\nabla^2 h^{\mu\nu}\right)
\nabla^0h_{\mu\nu}\right\}-\frac{I_2}{T}\, ,
\end{equation}
where $T$ is the interval of time in the integrations. 
In order to obtain the expression in (\ref{Hamiltonian}), we have used the partial 
integrations with respect to the time coordinate. 
Substituting the equations for massless graviton and massive one, we find that 
their energies are given by
\bea
\label{energy}
&& E_{m}=-\frac{1}{2\kappa^2T} \left( 1-\frac{2}{3}\alpha\Lambda+2b\gamma_\pm \right)
\int d^4x \sqrt{-g} \nabla^0h_{(m)}^{\mu\nu}\dot h^{(m)}_{\mu\nu}\, ,\\
&& E_{M}=\frac{1}{2\kappa^2T}(1-\frac{2}{3}\alpha\Lambda+2b\gamma_\pm)
\int d^4x \sqrt{-g} \nabla^0h_{(M)}^{\mu\nu}\dot h^{(M)}_{\mu\nu}\, .
\eea
In the pure Einstein gravity, where $\alpha=b=\gamma_\pm=0$, $E_{m}$ is positive, 
and therefore the integration
$\int d^4x \sqrt{-g} \nabla^0h_{(m)}^{\mu\nu}\dot h^{(m)}_{\mu\nu}$ is negative. 
Thus, at the non-critical point, $E_{M}$ is negative although both of the expressions 
for the energies $E_{m}$ and $E_{M}$ vanish at the critical point (\ref{3-19}).

We also consider the energy of the logarithmic mode. 
The energy of the logarithmic mode has the following expression:  
\begin{equation}
E_\mathrm{log}=-\frac{\alpha}{\kappa^2T}
\int d^4x \sqrt{-g} \nabla^0 h_{\mu\nu}^\mathrm{log}\dot h^{(m)}_{\mu\nu}\, ,
\end{equation}
which is positive but the mixed states of the massless and logarithmic mode can have 
negative energy \cite{2}. 
Thus we have to truncate the logarithmic mode by imposing the appropriate boundary 
condition in order to preserve the unitarity.

We now define and compute the total energy-momentum tensor and the conserved charge, 
which is called the AD mass. 
First, we define the energy-momentum tensor by the variation of the action 
with respect ot the metric, $g_{\mu\nu}+h_{\mu\nu}$, and separating the equation (\ref{3}) 
into a sum of the linear part of $h_{\mu\nu}$ and other parts. 
Then the energy-momentum tensor is given by
\be
\label{4-1}
T_{\mu\nu} = \left(1+2b\gamma-\frac{4\alpha\Lambda}{3}\right)
\tilde{\mathcal{G}}_{\mu\nu}^L+\frac{\alpha}{3} \left(g_{\mu\nu}\nabla^2
 -\nabla_\mu\nabla_\nu+g_{\mu\nu}\Lambda \right)R^L 
+\alpha \left( \nabla^2\tilde{\mathcal{G}}_{\mu\nu}^L-\frac{2\Lambda}{3} R^Lg_{\mu\nu} 
\right)\, .
\ee
We can confirm that $T_{\mu\nu}$ is covariantly conserved since the (RHS) of (\ref{4-1}) 
is covariantly conserved \cite{3,4}. Then we obtain the following expression of the AD mass,
\begin{equation}
\label{4-3}
M_\mathrm{AD}=r_0 \left( 1+2b\gamma-\frac{2\alpha\Lambda}{3} \right)\, ,
\end{equation}
where $r_0$ is the Schwarzschild radius. 
The first term in (\ref{4-3}) is contribution from the Einstein-Hilbert action, 
and the second terms those from the scalar terms in the action. 
At the critical point (\ref{3-19}), the AD mass vanishes, 
\begin{equation}
M^\mathrm{crit}_\mathrm{AD}=0\, .
\end{equation}
In the Minkowski background, the AD mass is obtained by choosing 
$\Lambda=0$ in (\ref{4-3}),
\begin{equation}
\tilde M_\mathrm{AD}=r_0(1+2b \gamma)\, .
\end{equation}
At the critical point (\ref{critical2}), $\tilde M_\mathrm{AD}$ also vanish.

\section{Conclusions}

In this paper, we have considered the four dimensional model whose action 
is given by the sum of the Einstein-Hilbert action and 
the terms including the square of the Weyl tensor coupled with a scalar field. 
In the action, in addition to the (A)dS background solutions, 
the Minkowski background is obtained as a vacuum solution. 
Note that the MInkowski background solution does not appear without including 
the scalar field. 
In the backgrounds, we have shown that the critical point appears 
by a special choice of the parameters. 
At the critical point, the non-minimal coupling $\gamma$ is negative in any backgrounds.

The energies of the propagating modes and the AD masses are obtained in the same way as in \cite{1}. 
The obtained energy of the massive mode is negative at the non-critical point and it vanishes 
at the critical point. 
The logarithmic mode appears at the critical point when the mass of the massive mode vanishes and 
the energy of the logarithmic mode is positive. 
The mode of mixed states of the massless and massive modes, however, have negative energy. 
In order to preserve the unitarity, the logarithmic mode should be truncated by an appropriate 
boundary condition.
The AD masses vanish at the critical point. 
We also investigated whether there is a scalar field or not 
and we find the energies and the AD mass vanish at the critical point.

It is interesting that our approach may be generalized for the case when 
scalar field is dimensionless and here it includes higher derivatives too.
Such theory has been proposed in Ref.~\cite{Elizalde:1994nz} 
as renormalizable quantum 
gravity with higher derivative scalar. Moreover, it has been demonstrated 
in Ref.~\cite{Weinberg:2008hq} that it can be very interesting 
effective proposal for inflation.

\section*{Acknowledgments}

We are grateful to S.~D.~Odintsov for the discussion when he stayed in 
Nagoya University. 
S.N. is supported by Global COE Program of Nagoya University (G07)
provided by the Ministry of Education, Culture, Sports, Science \&
Technology and by the JSPS Grant-in-Aid for Scientific Research (S) \# 22224003
and (C) \# 23540296.

\end{document}